\documentclass[9pt,twocolumn,twoside]{pnas-new-no-watermark}
\templatetype{pnasresearcharticle-v1}
\usepackage{mathtools}
\pdfoutput=1

\newcommand{\Pca}{\mathcal{P}}

\title{The more you test, the more you find: Smallest P-values become increasingly enriched with real findings as more tests are conducted}

\author[a]{Olga A. Vsevolozhskaya}
\author[b]{Chia-Ling Kuo}
\author[e]{Gabriel Ruiz}
\author[c]{Luda Diatchenko}
\author[d,1]{Dmitri V. Zaykin}

\affil[a]{Department of Biostatistics, University of Kentucky, Lexington, USA}
\affil[b]{Department of Community Medicine and Heath Care, Institute for Systems Genomics, University of Connecticut Health Center, Farmington, USA}
\affil[c]{The Alan Edwards Centre for Research on Pain, McGill University, Montreal, Canada}
\affil[d]{National Institute of Environmental Health Sciences, National Institutes of Health, USA}
\affil[e]{Summer Internship Program, National Institute of Environmental Health Sciences, National Institutes of Health, USA}

\leadauthor{Vsevolozhskaya} 

\significancestatement{Misinterpretation and abuse of significance testing and P-values have been put in the center of the scientific research controversy as major culprits responsible for poor replicability of studies. Multiple testing as a form of `P-hacking' is believed to promote false findings that fail to replicate in subsequent investigations. We show that paradoxically, massive multiple testing, exemplified by large scale genetic association studies, does not promote spurious findings. The more tests are performed, the higher the proportion of real findings is expected in a set of the smallest P-values of a predefined size. This remains the case even among those studies where no P-values reached statistical significance.}

\authordeclaration{The authors declare no conflict of interest here.}
\correspondingauthor{\textsuperscript{1} Corresponding author (e-mail: dmitri.zaykin@nih.gov)}

\begin{abstract} %250 words max
Increasing accessibility of data to researchers makes it possible to conduct massive amounts of statistical testing. Rather than follow a carefully crafted set of scientific hypotheses with statistical analysis, researchers can now test many possible relations and let P-values or other statistical summaries generate hypotheses for them. Genetic epidemiology field is an illustrative case in this paradigm shift. Driven by technological advances, testing a handful of genetic variants in relation to a health outcome has been abandoned in favor of agnostic screening of the entire genome, followed by selection of top hits, e.g., by selection of genetic variants with the smallest association P-values. At the same time, nearly total lack of replication of claimed associations that has been shaming the field turned to a flow of reports whose findings have been robustly replicating. Researchers may have adopted better statistical practices by learning from past failures, but we suggest that a steep increase in the amount of statistical testing itself is an important factor. Regardless of whether statistical significance has been reached, an increased number of tested hypotheses leads to enrichment of smallest P-values with genuine associations. In this study, we quantify how the expected proportion of genuine signals (EPGS) among top hits changes with an increasing number of tests. When the rate of occurrence of genuine signals does not decrease too sharply to zero as more tests are performed, the smallest P-values are increasingly more likely to represent genuine associations in studies with more tests.
\end{abstract}

\begin{document}

% \verticaladjustment{-2pt}

\maketitle

\thispagestyle{firststyle}

\ifthenelse{\boolean{shortarticle}}{\ifthenelse{\boolean{singlecolumn}}{\abscontentformatted}{\abscontent}}{}

% \section*{Introduction}

\dropcap{T}he scientific community is growing increasingly concerned with low replicability of research findings. Difficulty in replicating results of published research in subsequent investigations is attributed to a number of factors, including failure to follow good research practices or, in rare cases, outright fraud\cite{begley2015reproducibility}. However, with the proliferation of large, complex datasets, such as next generation sequencing data, misuse of statistical methods is put in the center of the controversy (e.g., refs. \cite{wasserstein2016asa} and \cite{greenland2016statistical}). Publication of a novel scientific result backed by statistical analysis has been traditionally accompanied by a P-value, a commonly used measure of statistical significance. The pressure to publish or perish coupled with a requirement for results to be significant may encourage researchers to try out various analyses of data, test multiple hypotheses and then report those P-values that reached statistical significance -- a part of the phenomenon coined as P-hacking\cite{simonsohn2014p}. As a result, P-values have come under fire and non-transparency of multiple testing has become associated with promotion of false findings  \cite{johnson2013revised, nuzzo2014statistical, halsey2015fickle, lazzeroni2014p, lai2012subjective, cumming2008replication}. 

A field where extensive multiple testing is common is genetic epidemiology. With genetic studies that use modern high-throughput technologies, millions of tests are performed in an agnostic manner in search of genetic variants that may be associated with a phenotype of interest. Only the best results with the smallest P-values are reported. While few genetic variants in genome-wide association studies (GWAS) may reach a strict genome-wide significance threshold (e.g., P-value $\leq$ $5\times 10^{-8}$) and can be considered to be reliable associations, many would not reach significance or fall on a borderline. Investigators have to rely on the set of the smallest P-values to decide which genetic variants are worthy of further investigation. Variability of P-values in replication studies and their inadequacy as predictors of future performance have been questioned \cite{lazzeroni2014p,halsey2015fickle}, making these decisions even more challenging. Furthermore, adhering to even more stringent genome-wide significance thresholds than those currently in use to safeguard against lack of replicability may increase needed sample sizes to impractical levels and gain in power may be counterbalanced by a potential decrease in quality of phenotypic measurements.

Interestingly, despite shortcomings of P-values as measures of support for a research hypothesis, a high fraction of the  borderline genetic associations had been reliably replicated \cite{panagiotou2012should}. This has been attributed to adoption of replication practices and improvement of statistical standards, including stringent significance thresholds\cite{ioannidis2011false}. Here we suggest that another important factor is a drastic increase in the number of tests in a single study compared to the pre-GWAS era, which leads to quantifiable enrichment of the smallest set of P-values in an experiment by genuine signals. Our analysis gives statistical support to the argument that it is illogical to pay a higher penalty for exploring additional, potentially meaningful relations in a study\cite{rothman1990no}.

To conceptualize our problem, envision a quantile-quantile (QQ) plot of P-values from a genetic association study. When there is an excess of small P-values, compared to what would be expected if none of the studied genetic variants had any effect on the outcome, such plot on a log scale would have a hockey stick shape, with a set of the smallest P-values deviating from the $45^\circ$ line. In actuality, some of these smallest P-values correspond to genuine signals, which we would color in red, and others to false signals, which we would color in blue. If we were privy to the information which effects are genuine and had access to multiple QQ plots from many different genetic studies, we could focus on colors of a single P-value, e.g., the minimum P-value (minP) in each study. Some of the minP's would turn out red, others blue, and the average would be purple. The depth of this purple color would represent the frequency that the minimum P-value corresponds to a genuine signal, taken across studies. In a particular study, the probability that a signal is genuine can be estimated using the Bayesian approach. In this approach, one needs to provide the prior information, the external knowledge, such as the chances that a randomly selected genetic variant is genuine. After observing data, the posterior probability that a given minP corresponds to a genuine signal can be determined. This probability would reflect the degree of assurance and correspond to a shade of purple anywhere between blue and red, although in reality the color is either red or blue. Despite this uncertainty due to estimation based on data, the average of these posterior probabilities over multiple genetic studies would turn out to be exactly the same as the true degree of the purple shade, provided the prior information is correctly specified. This thought experiment is the background of the model that we evaluate here in relation to the number of statistical tests in similarly powered studies that search for associations of multiple potential predictors with an outcome. We find that on average, a manageable set of the smallest P-values becomes steadily saturated with genuine effects as more testing is done in every study, even among those studies where none of the P-values cleared a multiple testing adjusted significance threshold.

\section*{Results}
It is illustrative to describe our model in terms of a genome-wide association scan, where single nucleotide polymorphisms (SNPs) or more generally, alleles of genetic loci, carry signals that reflect the strength of association with an outcome, for example, susceptibility to disease. Various effect sizes across the genome occur with different frequencies, in other words, the SNP-specific effect size (magnitude of a signal) across the genome forms a distribution. We may consider testing which SNPs have an effect size that is at least $\gamma^0$ in magnitude, and define the null hypothesis $H_0$ that a particular effect size is smaller than $\gamma^0$ with the alternative $H_A$, that it is larger than $\gamma^0$.

In this representation, effect sizes of all SNPs can be divided into two sets, the null set $\Gamma_0$, and the set $\Gamma_A$ with the effects of SNPs in that set that are larger than $\gamma^0$. The proportion of SNPs that fall into $\Gamma_0$ can be regarded as the prior probability of $H_0$, $\pi = \Pr(H_0)$; then $\Pr(H_A) = 1-\pi$.

We consider P-values derived from commonly used test statistics, such as chi-squared, $F$, normal $z$, and Student's $t$ statistics. In two-sided versions of these statistics, P-values are for the test $\gamma^0=0$, i.e., the effect size is assumed to be exactly zero under $H_0$ (point null hypothesis). Even though the computation of P-value may be carried out under the point $H_0$, we can still define the sets $\Gamma_0$ and $\Gamma_A$ without the point null assumption ($\gamma^0=0$) and ask the question whether the originating, actual effect size falls into $\Gamma_A$ and thus, by our definition, is genuine. Having obtained a particular P-value, we can evaluate the probability of a signal to be genuine as the posterior probability $\Pr(H_A \mid \text{P-value})$=$\Pr(\gamma \in \Gamma_A \mid \text{P-value})$ via the Bayesian approach (e.g. ref. \cite{wakefield2007bayesian}). This requires prior information, which in our model is summarized by the distribution of the effect size $\gamma$ across all SNPs in the genome. The computation does not change by the fact that the minimum P-value is selected out of all P-values in a genome-wide scan: the probability $\Pr(H_A \mid \text{minimum P-value})$ is computed the same way as for a random P-value that was not subject to selection. This highlights resistance of the Bayesian approach to selection bias \cite{dawid1994selection,senn2008note}. If the effect size distribution had been known precisely, these posterior probabilities would have been exact in the sense that the average $\Pr(H_A \mid \text{minimum P-value})$ taken across a large number of additional studies from the same population and with the same set of SNPs would yield a correct chance that the minimum P-value had originated from a genuine signal. Thus, we can talk about the average of such computed probabilities that the minimum P-value stemmed from a genuine signal. This average (expectation), taken across many replication studies is also equal to the expected proportion of genuine signals (EPGS), due to dichotomization of the effect size distribution into the null and the alternative groups, $\Gamma_0$ and $\Gamma_A$. Extending this from the minimum P-value to the top $u$ smallest P-values, we define EPGS as the expected proportion of genuine signals among $u$ smallest out of $k$ total sorted list of P-values, $\{p_{(1)}, \ldots p_{(k)}\}$. We refer to the minimum P-value, $p_{(1)}$ as minP.

As noted above, the posterior probability that a signal with the minimum P-value is genuine, $\Pr(H_A \mid \text{minP})$, does not depend on the number of tests, $k$, and does not require a correction for multiple testing. If one had access to a large number $B$ of independent studies and took the minP from each, then the empirical estimate of EPGS would simply be the average:
\begin{eqnarray}
   \text{EPGS} = E\{\Pr(H_A \mid \text{minP})\}
  \approx \frac{1}{B} \sum_{i=0}^B \Pr(H_A \mid \text{minP}_i) \label{bayesepgs}
\end{eqnarray}
EPGS can also be evaluated analytically as shown in Section S1 of \textit{SI Text}. As expected, the theoretical form of EPGS shows dependency on marginal distributions of effect sizes that absorb the prior information used in the Bayesian approach. It also reveals dependency of EPGS on the number tests. A seeming paradox is that the average of posterior probabilities in Eq. \ref{bayesepgs} can be computed without knowing the number of tests in any of $B$ studies and yet the average in its theoretical form depends on the number of tests. An intuition for this can be gained by looking at the posterior probability as an expectation, i.e., the average of an indicator,  $E\left\{I(\gamma \in \Gamma_A) \mid \text{minP}\right\}$. If a random P-value was substituted in place of the minP, then by the rule of iterated expectation EPGS would be equal to the expected value of the prior distribution. However, the distribution of P-values is modified due to selection in a way that depends on the number of tests, $k$. EPGS would similarly depend on $k$ if it was evaluated over a threshold $\alpha$, corrected for multiple testing as $\alpha/k$. 

\begin{figure}
\centering
\includegraphics[width=0.4\textwidth]{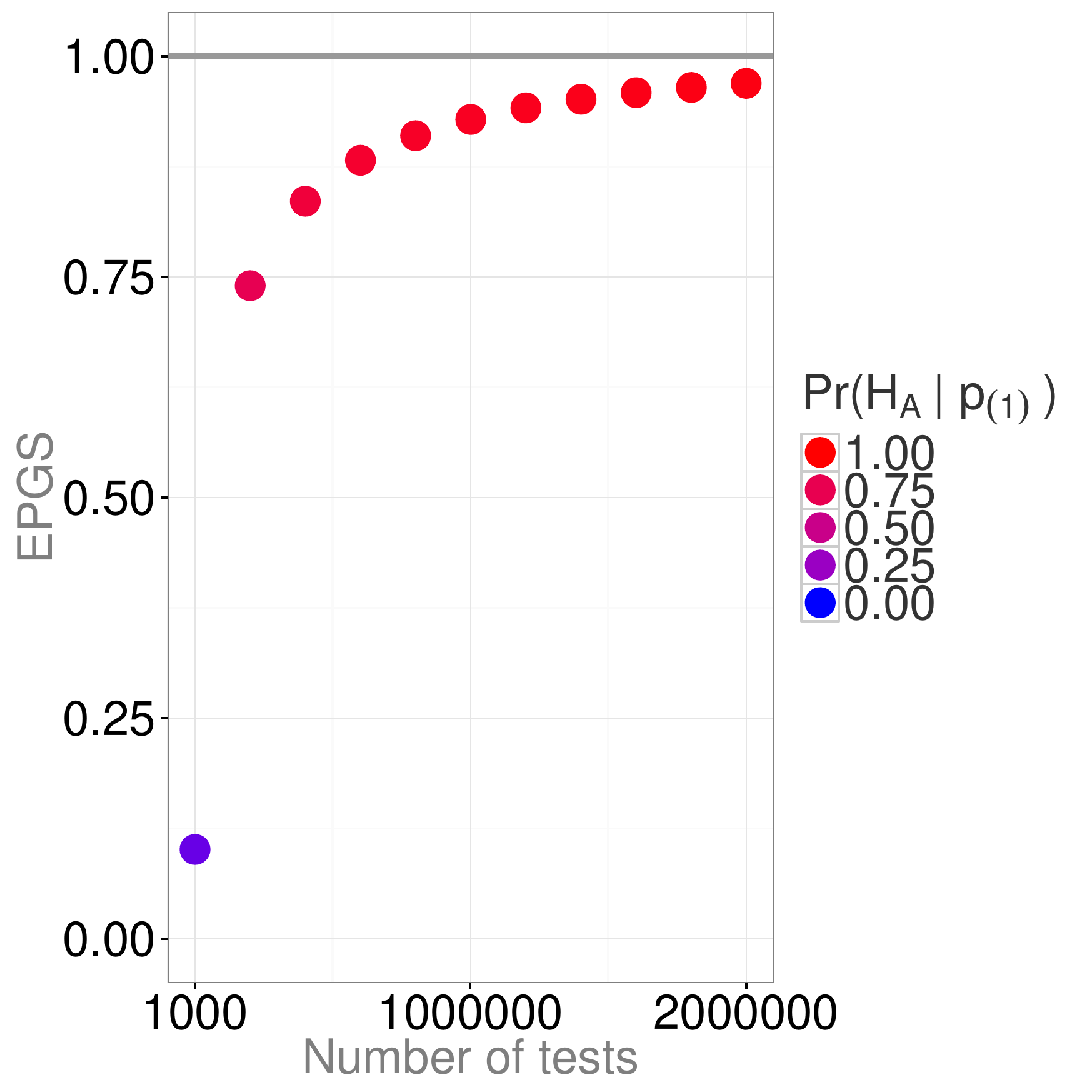}
\caption{Proportion of genuine signals versus the number of tests under the assumption of constant $\Pr(H_A)$. This plot depicts how the chances that the smallest P-value is a true finding converges to 1 with increased number of tests.}
\label{fig1}
\end{figure}

Our goal is  to evaluate the behavior of EPGS as the number of tests, $k$, increases. We first consider changes in the probability that the smallest P-value corresponds to a true signal as more tests are performed, i.e., we consider changes in $\Pr(H_A \mid \text{minP})$ as the number of tests, $k$, approaches infinity. Assuming that the prior probability of the alternative hypothesis, $\Pr(H_A)$, is constant, \textit{SI Text} shows that $\Pr(H_A \mid \text{minP}) \rightarrow 1$ as $k \rightarrow \infty$.  The assumption of constant $\Pr(H_A)$ is not necessarily unreasonable with high-throughput sequencing data because the chances of finding a causal variant in a discovery study do not necessarily decrease if a larger number of variants is tested. Therefore, our theoretical findings justify a carry-forward of the most-significant association variant from discovery to replication stage. To validate analytical results, we conducted a simulation study and calculated empirical values of EPGS for a different number of tests. To compute the approximate EPGS (Section S2, \textit{SI Text}), it was assumed that the prior probability of a true finding is 0.1\% and that the expected value of a $\chi^2$ statistic with one-degree-of-freedom is 5 -- a low ``typical'' test statistic value for a genomewide association scan. Next, in Fig. \ref{fig1}, we plotted the values of EPGS against the number of tests, ranging from 1,000 to 2 million.

Fig. \ref{fig1} illustrates convergence of the probability for the strongest signal to be a true finding to one as more tests are performed. The results are shown for the smallest P-value, but we can also evaluate the degree of enrichment by genuine signals for the next ordered P-values (second, third, and so on).

\begin{figure*}
\centering
\includegraphics[width=0.3\textwidth]{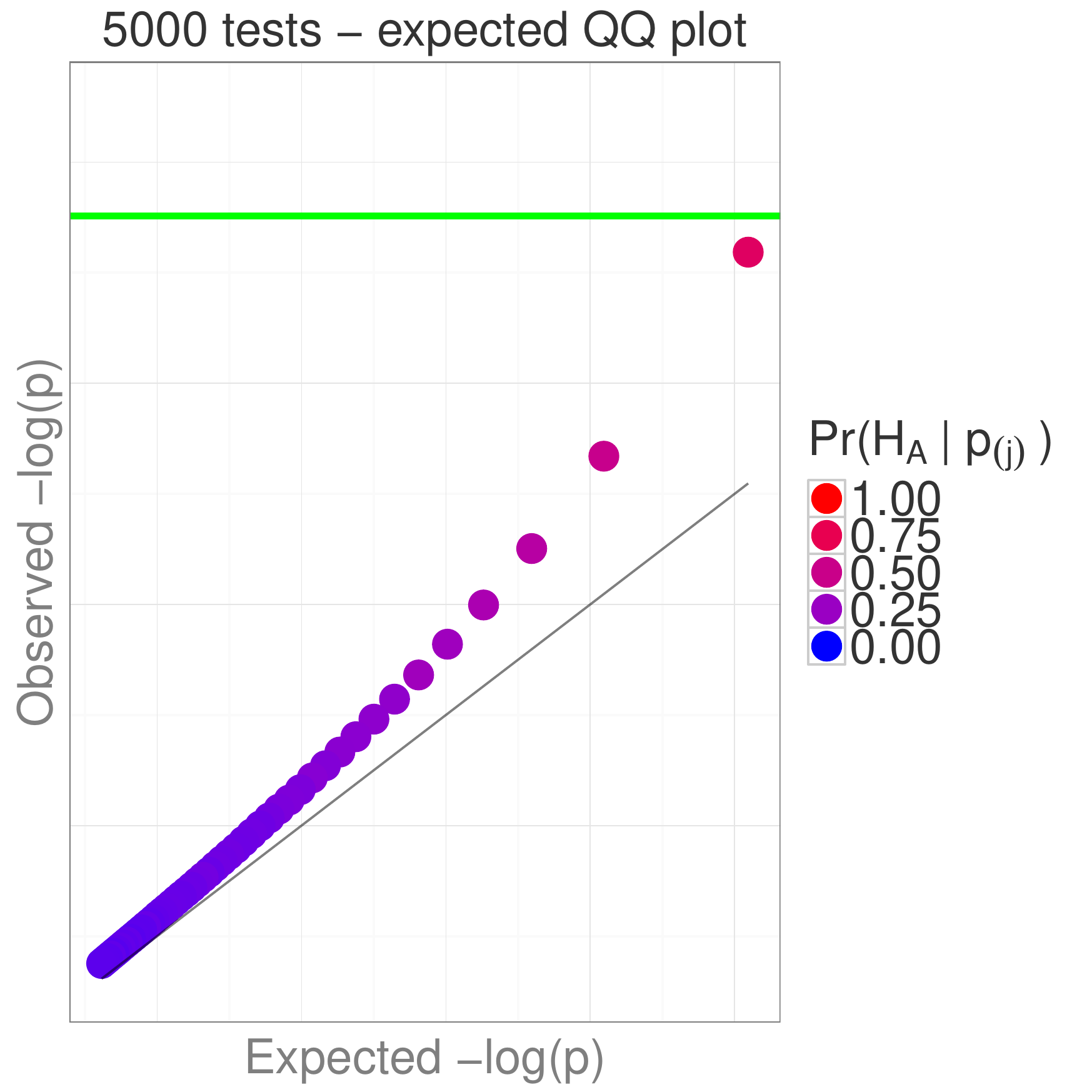}
\includegraphics[width=0.3\textwidth]{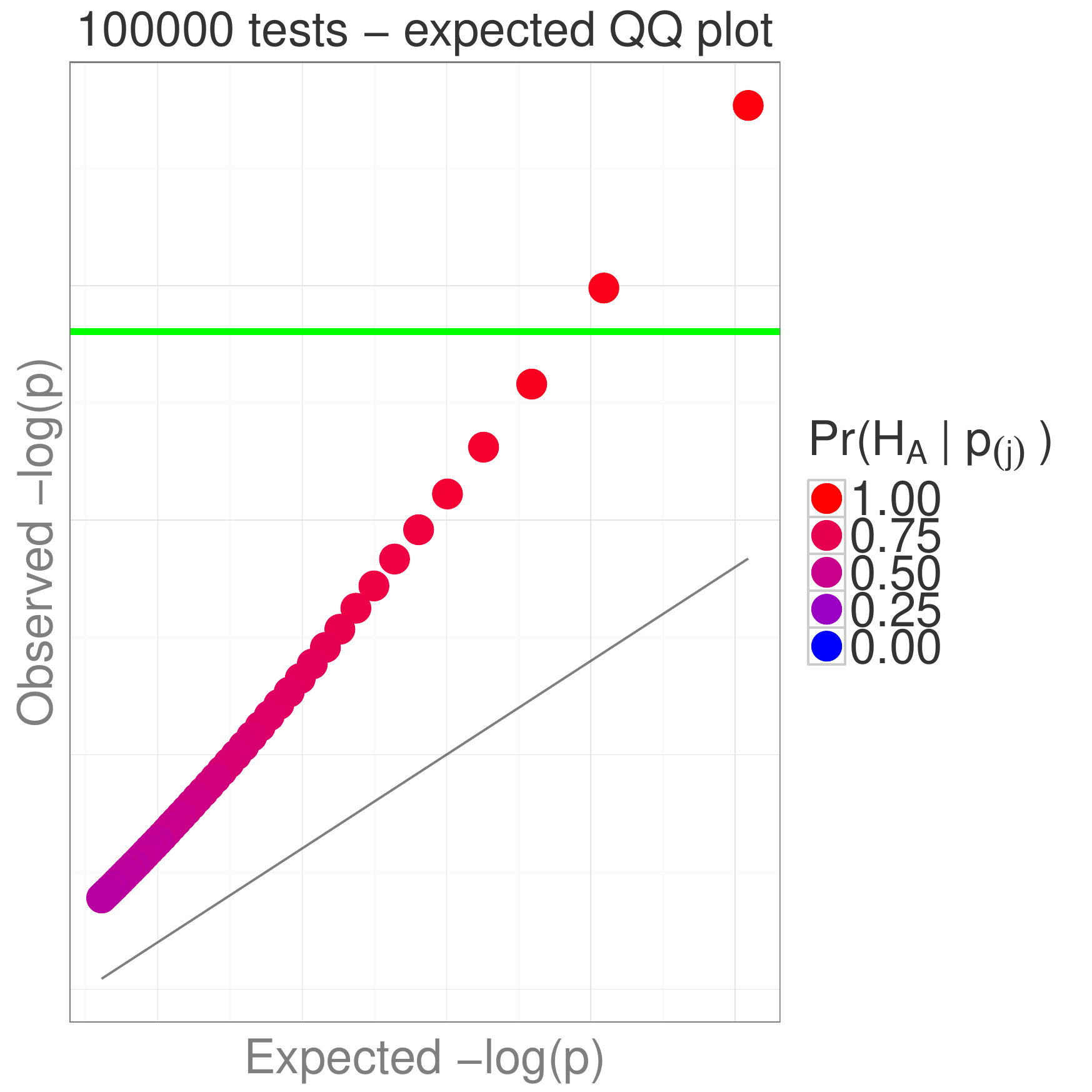} 
\includegraphics[width=0.3\textwidth]{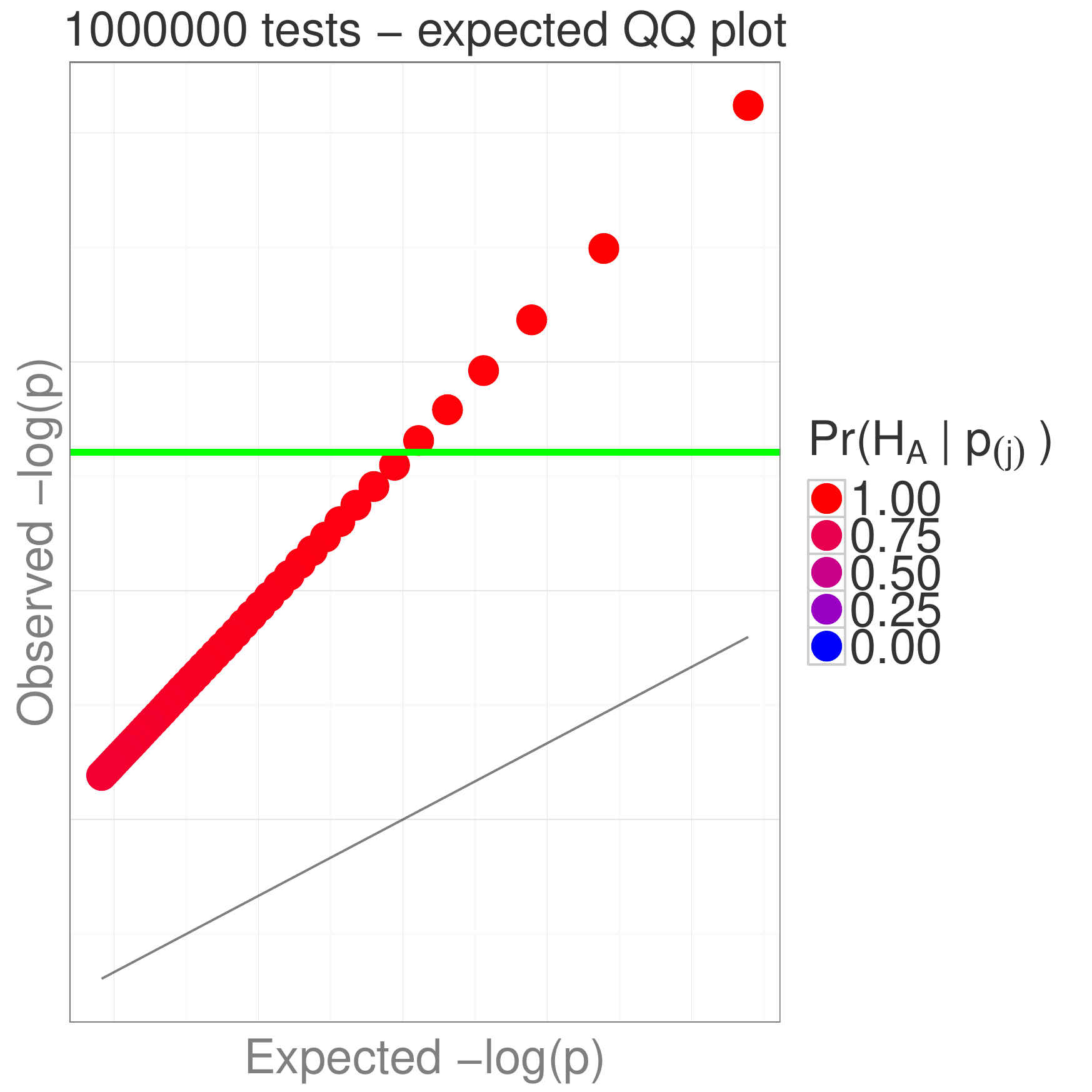} 
\caption{Proportion of genuine signals among top 50 P-values with the prior probability of a true finding of 2\%. The color of the dots represents the posterior probability of a genuine result, ranging from blue ($\Pr(H_A \mid p) = 0$) to red ($\Pr(H_A \mid p) = 1$). The green line indicates Bonferroni-adjusted significance threshold.}
\label{fig2}
\end{figure*}

\begin{figure*}
\centering
\includegraphics[width=0.3\textwidth]{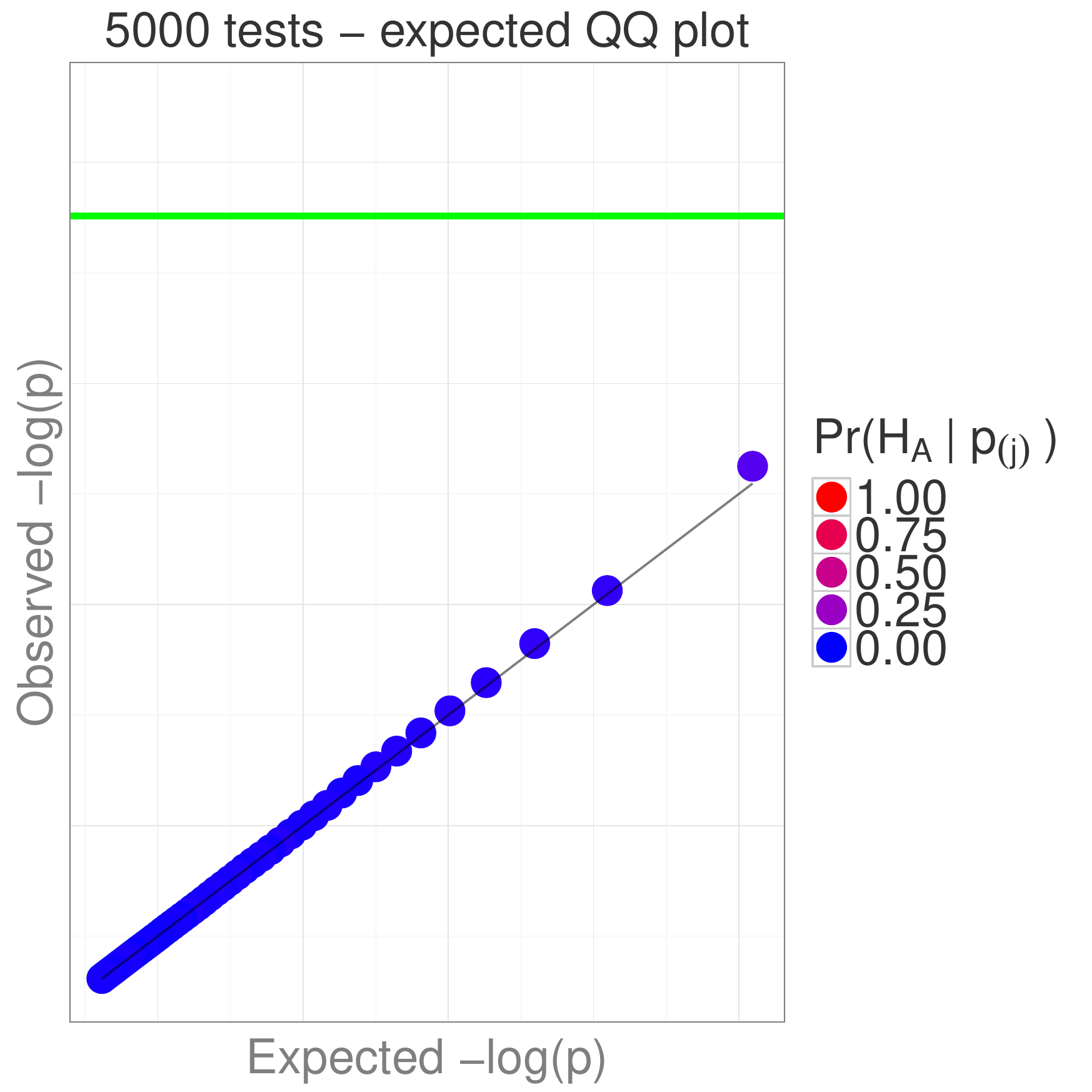}
\includegraphics[width=0.3\textwidth]{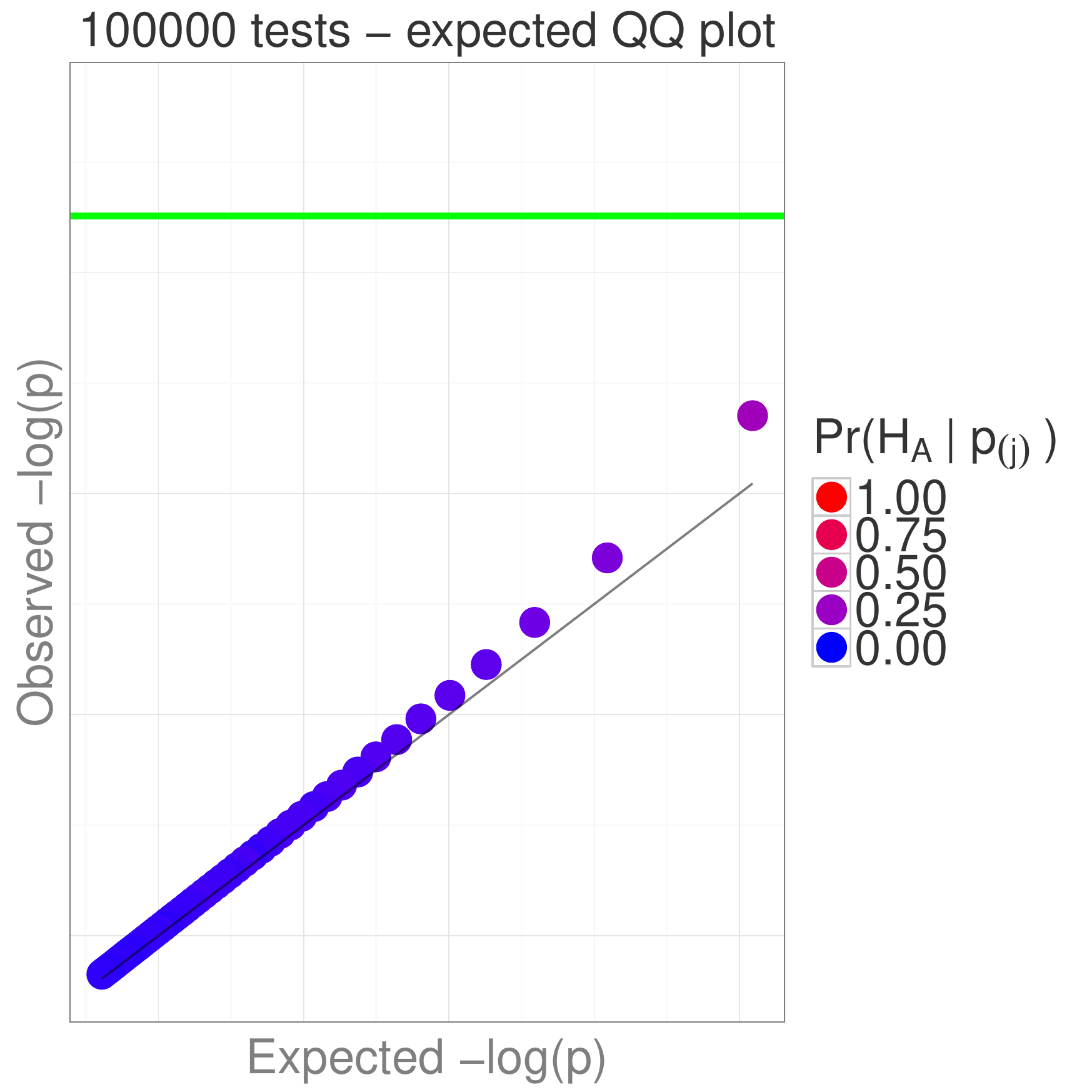} 
\includegraphics[width=0.3\textwidth]{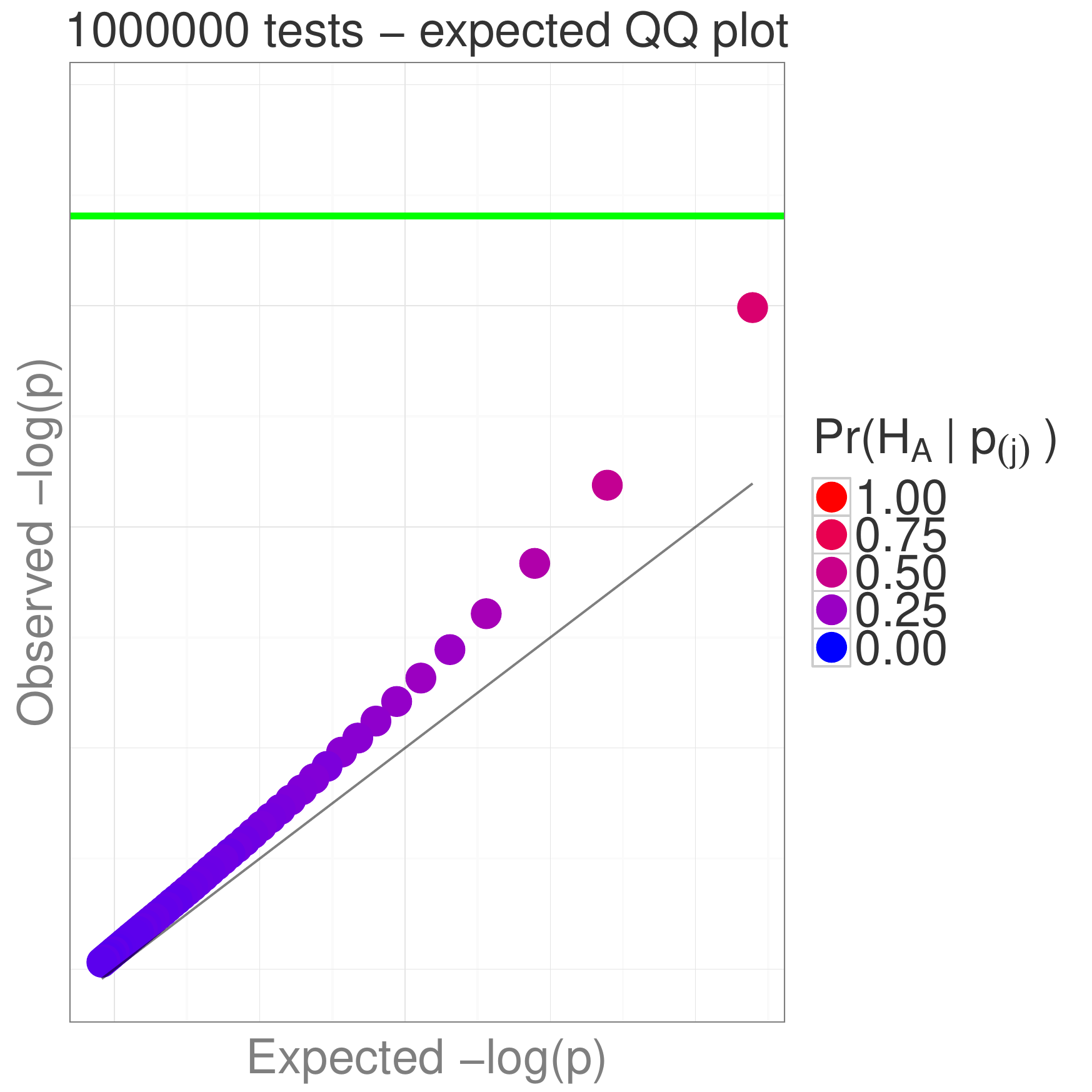} 
\caption{Proportion of genuine signals among top 50 P-values with the prior probability of a true finding of 0.1\%. The color of the dots represents the posterior probability of a genuine result, ranging from blue ($\Pr(H_A \mid p) = 0$) to red ($\Pr(H_A \mid p) = 1$). The green line indicates Bonferroni-adjusted significance threshold.}
\label{fig2_new}
\end{figure*}

To determine how the chances of the top $j$th P-value to be a true finding vary with the number of tests, we calculated the empirical probabilities $\Pr(H_A \mid p_{(j)}), j=1, \ldots, 50$ for the top 50 P-values. For the effect size distribution, $\gamma \sim \Gamma(\gamma)$, we assumed an L-shaped distribution as suggested by the population genetics theory \cite{otto2000detecting}. Specifically, we assumed the Gamma(0.25, 5) distribution, and set the expected rate of true associations to either 2\% or 0.1\%. The results are summarized by a QQ P-values plot in Fig. \ref{fig2} and Fig. \ref{fig2_new}. Each panel of  Figures \ref{fig2}-\ref{fig2_new} shows the observed $-\log_{10}(p_{j}), j = 1, \ldots, 50$ versus the expected $-\log_{10}(p_{j})$ under the point $H_0$ for $k=500, 10,000$ and $1,000,000$ tests. The green line corresponds to the Bonferroni-corrected significance threshold at $5\%/k$ level. The color of the dots represents the empirical probability of a true positive result out of 1,500 simulations, ranging from blue ($\Pr(H_A \mid p) = 0$) to red ($\Pr(H_A \mid p) = 1$). Figures \ref{fig2} and  \ref{fig2_new} show enrichment of top P-values by true signals and illustrate how the rate of enrichment depends on the rate of occurrence of genuine signals.

\begin{figure*}
\centering
\includegraphics[width=0.3\textwidth]{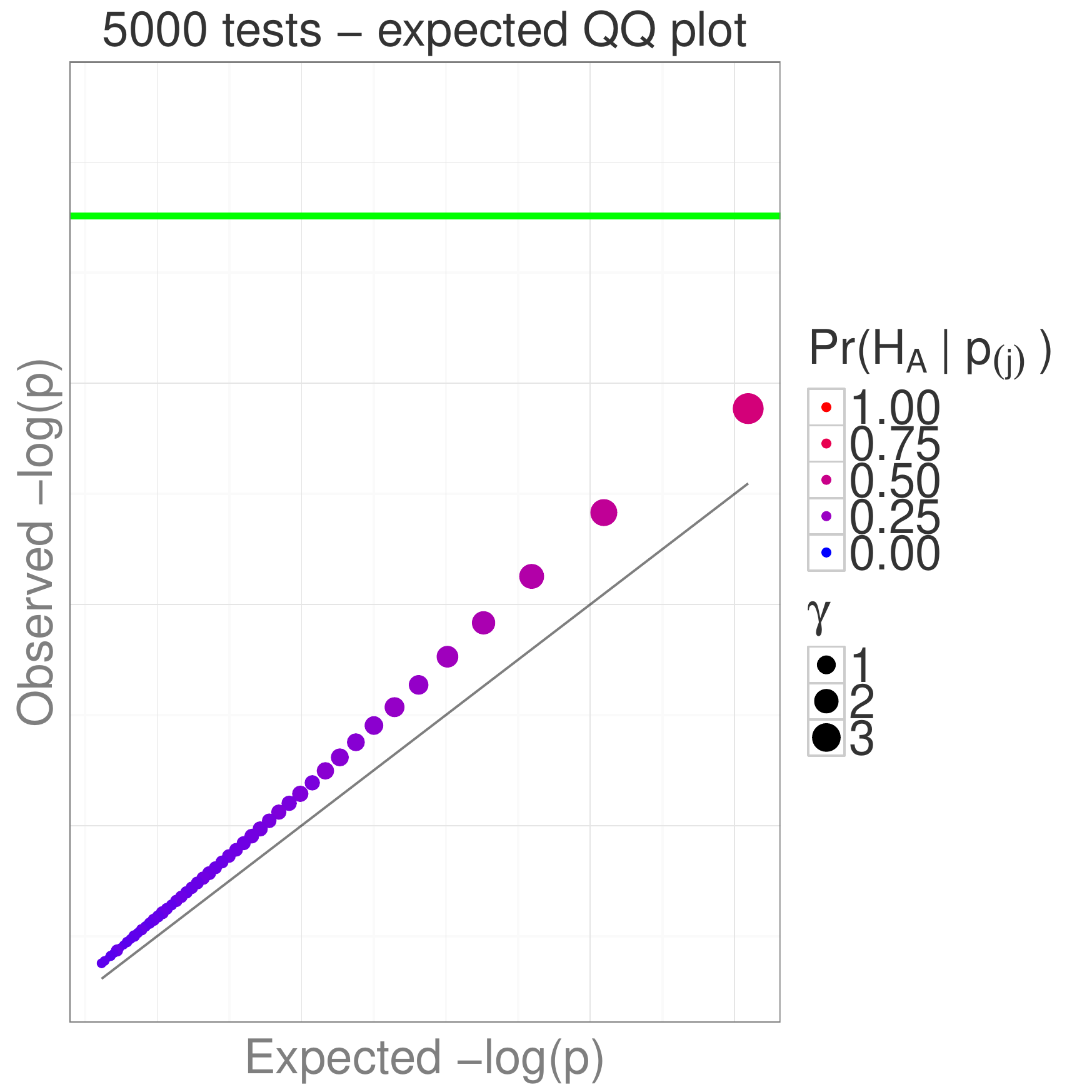}
\includegraphics[width=0.3\textwidth]{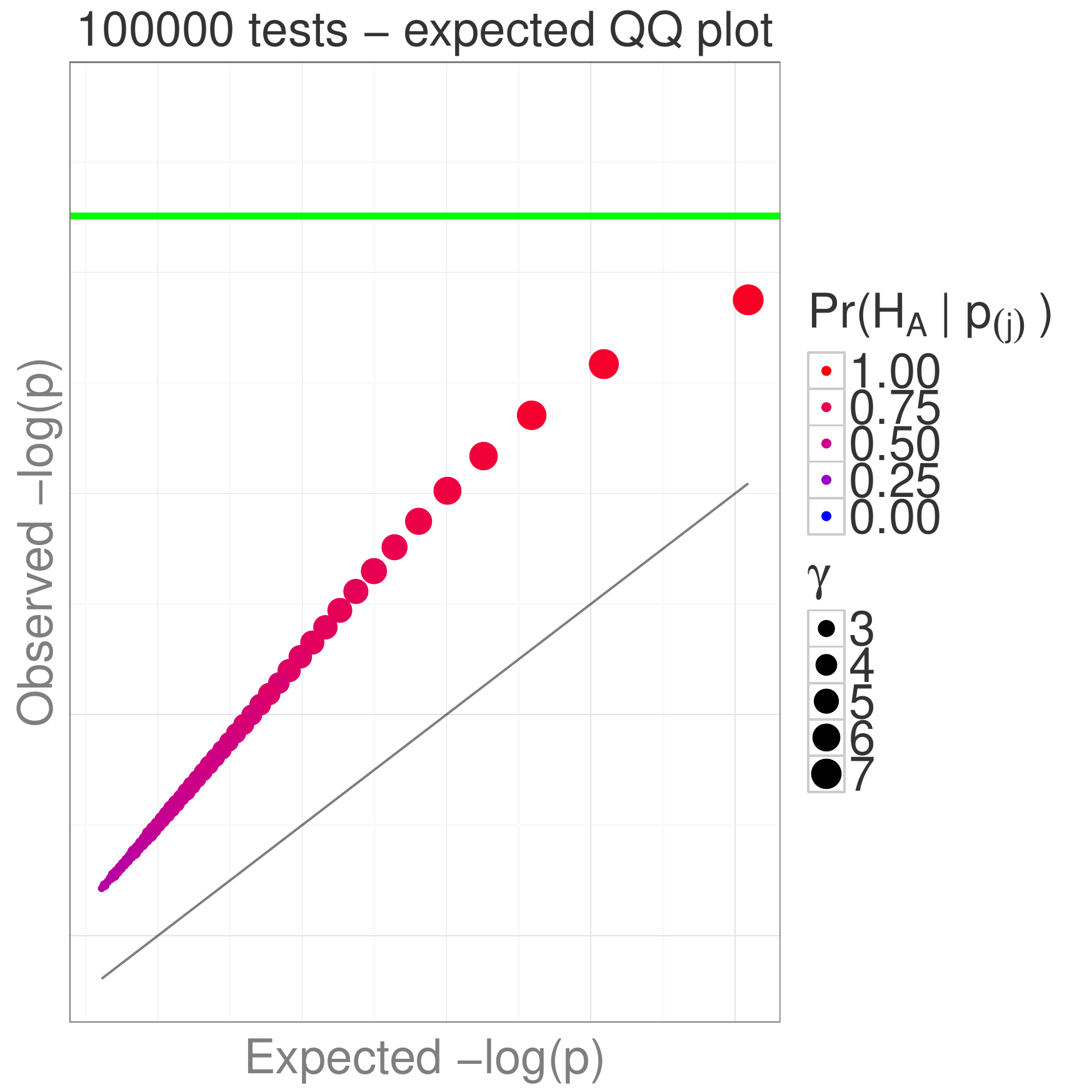}
\includegraphics[width=0.3\textwidth]{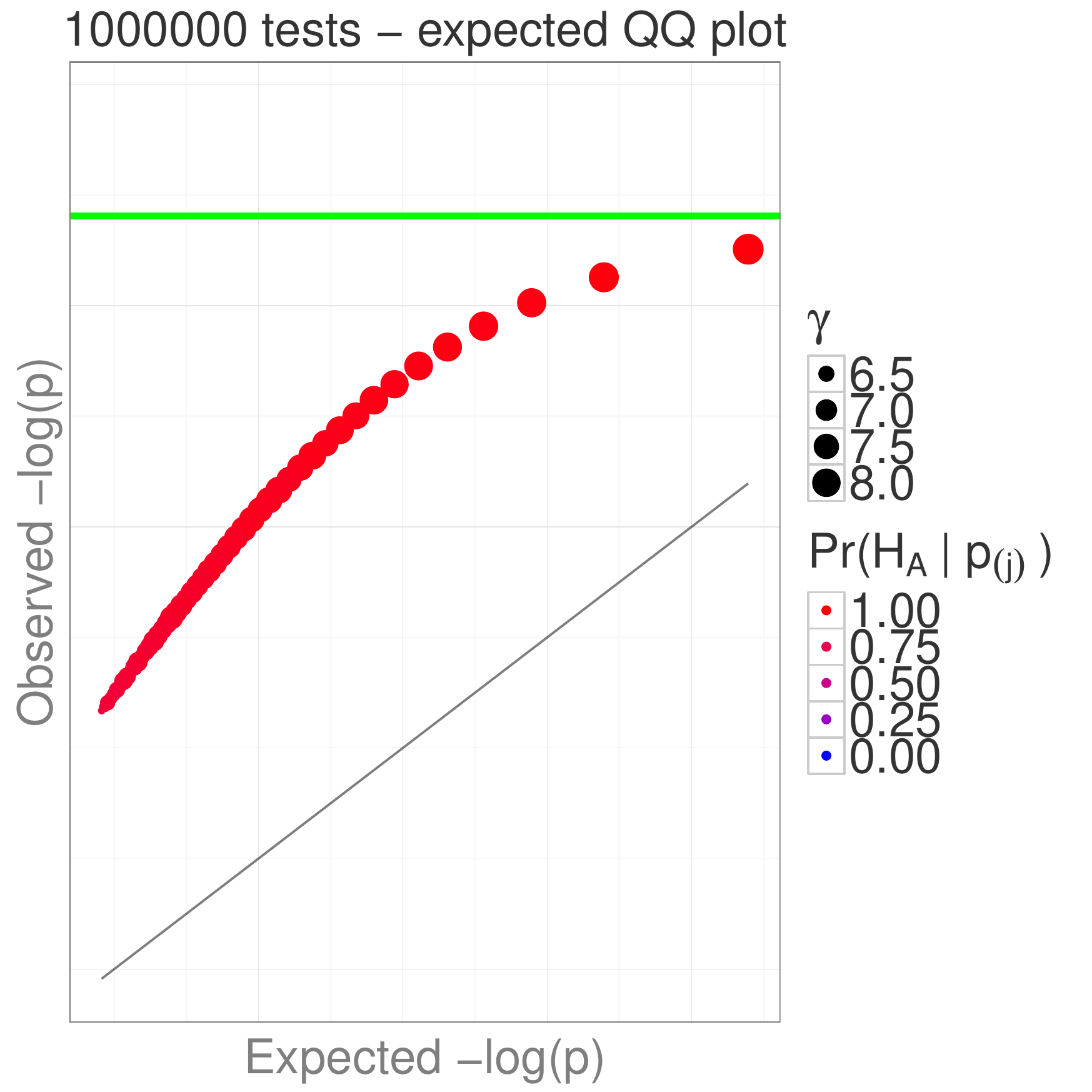}
\caption{Proportion of genuine signals among top 50 P-values, none of which passed a significance threshold ($\min(p) > 0.05/k$). The color of the dots represents the posterior probability of a genuine result, the size of the dots represents the magnitude of the effect size, and the green line indicates Bonferroni-adjusted significance threshold.}
\label{fig3}
\end{figure*}

It is evident from Figs. \ref{fig2}-\ref{fig2_new} that the probability of a P-value to be a true association is increasing even among P-values that did not reach $5\%/k$ significance level (note the change in the color of the dots below the green threshold line). Thus, we modified our simulations in a way in which experiments where the minimum P-value was significant were discarded. A practice to keep only those studies where no significances were found would clearly go against common sense. Yet, the enrichment of top hits by genuine signals still takes place: Fig. \ref{fig3} shows the results. Note that it is impossible in these graphs for the dots to cross the Bonferroni green line, because experiments with significances were discarded. Still, even among these experiments without any significant P-values, chances that top hits represent true findings increase with the number of tests. For instance, with 1 million tests and among top 50 P-values, none of which reached statistical significance, the empirical probability of a true association ranges from $\sim 35\%$ for the $p_{(50)}$ to $\sim 92\%$ for $p_{(1)}$. Moreover, the magnitude of the effect size, $\gamma$, is increasing with both the order of P-values and with the number of tests.

Figures \ref{fig1}-\ref{fig3} were constructed according to the traditional point null hypothesis assumption of zero effect size. However, more realistically, the null hypothesis may be represented by a set of non-zero, negligibly small effect sizes. For this scenario, we used effect size estimates from Park et al. \cite{park2010estimation}, who provided the number of susceptibility loci and the distribution of their effect sizes measured as a function of the odds ratio (OR) for breast, prostate and colorectal (BPC) cancers. Signals that correspond to $H_0$ were defined as those with the OR in the range of 1/1.01 to 1.01 with the prior probability of $H_0$ equal to one minus the estimated proportion of susceptibility loci for BPC cancers. Table \ref{tab1} summarizes the results and shows how the average odds ratio (OR) corresponding to the smallest P value changes with the increase in the number of tests. From Table \ref{tab1}, it is clear that even without the assumption of the point null hypothesis, there is an enrichment of true signals with high effect size magnitude among the top hits.
\begin{table}%[tbhp]
\centering
\caption{Comparison of the average odds ratios without the point null assumption.}
\begin{tabular}{lcc}
Number of tests & $\Pr(H_A \mid p_{(1)})$ & $\widehat{OR}$ \\
\midrule
$k = 100 $ & 0.0102 & 1.0105 \\
$k = 1,000$ & 0.0538 & 1.0164 \\
$k = 10,000$ & 0.2596 & 1.0454 \\
$k = 50,000$ & 0.6234 & 1.1101 \\
$k = 100,000$ & 0.8135 & 1.1507 \\
$k = 500,000$ & 0.9946 & 1.2235\\
\bottomrule
\end{tabular}
\label{tab1}
% \addtabletext{nomenclature for the TSs refers to the numbered species in the table.}
\end{table}

The enrichment of top hits with signals carrying increasingly large effect sizes, as the number of tests increases can be characterized analytically (Section S4 of \textit{SI Text}). Instead of making a distinction between effect sizes that are large enough to be considered genuine and correspond to $H_A$ and a set of smaller effect sizes that correspond to $H_0$ we can model all $k$ tested signals together as arising from an effect size distribution. Such distribution may be L-shaped and have a sizable spike around zero to reflect preponderance of signals carrying small effects. In terms of the genome wide association scans, such distribution reflects the prior knowledge that a randomly chosen SNP has a small effect size with high probability. Then the expected effect size for ordered P-values can serve as a measure of enrichment of top hits by genuine signals. This measure is not the same as the expected value of the observed (estimated) effect sizes for top hits, because the latter is a subject to selection bias and overestimates the actual average effect size. The approximation we derived (\textit{SI Text} Eq. \ref{siexpgamma}) estimates the expectation taken across top hits of random experiments and allows for experiments themselves to be subject to selection, as in  Fig. \ref{fig3}, where the average was taken only across nonsignificant experiments, i.e., across those where minP did not reach the Bonferroni threshold (minP$>0.05/k$). The approximation  reveals that the enrichment of top hits by genuine signals with increasingly large effect sizes is to be expected even among multiple testing experiments without statistically significant findings.

\begin{figure}
\centering
\includegraphics[width=0.4\textwidth]{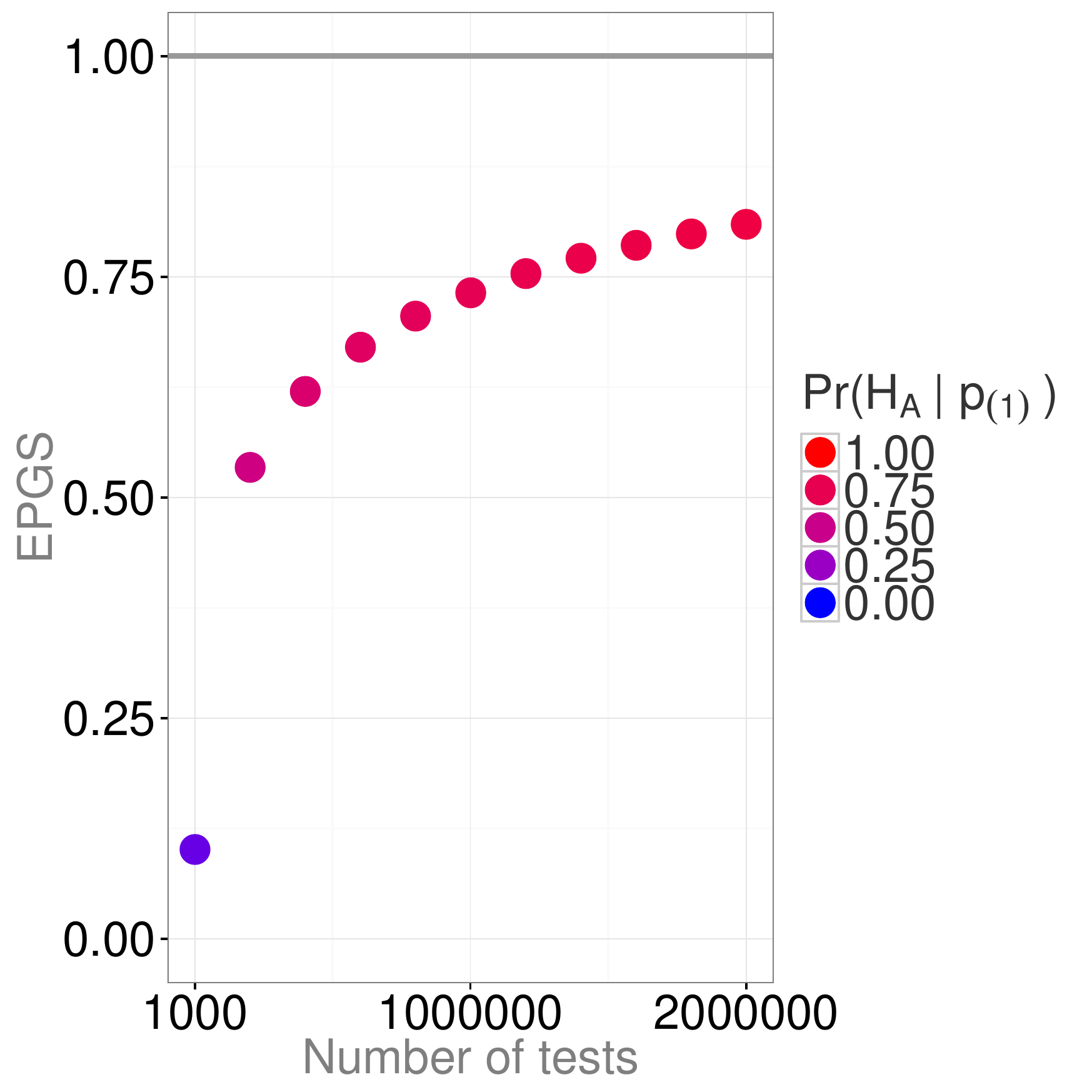}
\caption{Proportion of genuine signals versus the number of tests ($k$) under the assumption of $\Pr(H_A) = 0.01\% / \ln(k)$. This plot depicts how the chances of the smallest P-value to be a true finding still converge to 1 even if the rate of true findings vanishes with $k$.}
\label{fig4}
\end{figure}

The results displayed in Figures \ref{fig1} -\ref{fig3} and Table \ref{tab1} are for the condition of constant rate of encountering a true positive. The assumption of constant rate of $\Pr(H_A)$ regardless of the number of tests may not always be reasonable. When the rate decreases and eventually reaches a positive constant, EPGS will still increase to one, although at a slower pace compared to the starting rate. However, it is also possible for the of $\Pr(H_A)$ to approach zero slowly enough so that EPGS will still increase with $k$. These results are given in \textit{SI Text}, where we investigated the limiting behavior of EPGS when $\Pr(H_A)$ is a decreasing function of $k$, $h(k)$. In particular, if the probability of a genuine signal vanishes with $k$, i.e., $\lim_{k \to \infty}h(k) = 0$, but
\begin{eqnarray}
  \label{eq:hk}
  \lim_{k \to \infty}h(k)\cdot k = c,
\end{eqnarray}
where $c$ is a constant, then $\text{EPGS}\rightarrow 0$ as $k \rightarrow \infty$. However, when the rate of occurrence of genuine signals is slowly decreasing with $k$, EPGS may still increase, with one example being a logarithmic decrease, $h(k) = \Pr(H_A)/ \ln(k)$. As an illustration, we modified Fig. \ref{fig1} by assuming such decrease. The results are summarized in Fig. \ref{fig4} and show that EPGS is still converging to one but at a slower rate.

While our results are derived assuming independence of P-values, we showed earlier that local dependencies, such as expected in GWAS induce little bias in ranking probabilities (\textit{\textbf{SI Text}} Eq. \ref{rankprob}) through which the EPGS can be expressed. In particular, the distribution of the smallest P-values may retain the same form as under independence, but with the number of tests reduced from $k$ to a smaller number, $k_e$, an ``effective number of tests,'' due to local correlation \cite{kuo2011novel}.

\section*{Discussion}
It is hard to overstate importance of statistical analysis as a key factor affecting replicability of research. Relatively high rate of replicability seen in genetic association studies in comparison to the rest of epidemiology has been attributed to several factors, including increased awareness of researchers about the dangers of multiple testing and to the subsequent adoption of appropriately stringent significance thresholds. However, an overlooked factor may have been that selection of top smallest P-values out of many tests leads to enrichment of these top hits with genuine signals as the number of tests increases. 

Our basic model for this enrichment is when the rate of occurrence of genuine signals is constant. The rate of occurrence can be allowed to decrease with the number of tests for the enrichment to take place. However, the decrease should either be sufficiently slow, or should stabilize at some non-zero baseline level. In the latter case the rate becomes constant, although lower than the rate for the small number of tests.

While we are primarily motivated by genetic association studies, our model is general. As an illustration, imagine a completely uninformed epidemiologist studying effects of various predictors on susceptibility to a disease. The epidemiologist is oblivious to any external knowledge regarding possible effects of predictors on the outcome and simply tests every predictor in sight. In this scenario, the rate with which truly associated predictors are tested does not diminish as additional predictors are tested. 

At the end of the day, the predictor yielding the smallest P-value is reported as a potentially true association. This strategy is often perceived with disdain as `data torturing.' However, a predictor with the smallest P-value in such a study becomes increasingly less likely to be a spurious association as more tests are performed. Therefore, an epidemiologist that tested effects of one hundred random exposures on a disease and reported the smallest of one hundred P-values with no regard to its significance is more likely to be correct in identifying a truly associated effect than her colleague that tested only ten exposures. On the other hand, a knowledgeable epidemiologist would study most plausible predictors first and thus the rate of occurrence of genuine signals would drop with any additional testing, as a consequence. However, it is reasonable to suppose that as the prior, subject matter expertise is exhausted, she would be settled at a lower, yet a constant rate scenario, which is a basis of our model. 

Our observations regarding the enrichment of top hits with true positives model a sequential testing of potential predictors. A very different scenario would be testing for all possible higher-order interactions among predictors. In such scenario, the number of tests grow exponentially, and it is possible that the rate of genuine signals $\Pr(H_A)$ would quickly approach zero because of a steep increase in the number of tested combinations.

Our analysis reflects limitations of P-values as summary measures of effect size. Distribution of P-values for commonly used test statistics depends on the product of the sample size, ($N$ or $\sqrt{N}$) and a measure of effect size, $\delta$, scaled by the variance $\sigma^2$ (or $\sigma$). For example, when the outcome is a case/control classification and the predictor is also binary, the standardized effect size can be expressed in terms of the squared correlation:
\begin{eqnarray}
   \gamma &=& N \times R^2 = N\times (\delta/\sigma)^2 \nonumber \\
  &=& N \times \frac{(p_1-p_2)^2}{\tilde{p}(1-\tilde{p}) \left[w(1-w)\right]^{-1}}, \nonumber \\
\end{eqnarray}
or the log of the odds ratio:
\begin{eqnarray}
    \gamma &=& N \times (\delta/\sigma)^2 = N \times \frac{\left[\log(\text{OR})\right]^2}{\frac{1}{w} \frac{1}{p_1(1-p_1)} + \frac{1}{1-w} \frac{1}{p_2(1-p_2)}} \nonumber \\
     &\approx& N \times \frac{\left[\log(\text{OR})\right]^2}{ \left[\tilde{p}(1-\tilde{p}) w(1-w)\right]^{-1} },
\end{eqnarray}
where $p_1$,$p_2$ are frequencies of exposure in cases and controls respectively, $\tilde{p}$ is the pooled frequency of exposure, and $w$ is the proportion of cases. Scaling of $\delta$ by $\sigma$ itself limits interpretability, and the second major limitation is the conflation of this standardized measure of effect with $N$. Thus, the interpretation of our results is most straightforward for studies where P-values are derived from statistics with similar sample sizes, such as in genome wide association scans.

Theoretical results developed here allow researchers to quantify the expected enrichment of the smallest P-values by genuine signals as a function of the number of tests, given external information about the effect size distribution. Qualitatively, we demonstrate that the expected proportion of genuine associations among the smallest P-values of multiple testing experiments is expected to increase with the number of tests. Rothman made an argument based on implausibility of the global null hypothesis that scientists should not be reluctant to explore potentially wrong leads due to fear of penalty for peeking\cite{rothman1990no}. We make this argument more formal and suggest that more testing should be preferred to less.

\section*{Acknowledgements}
This research was supported in part by the Intramural Research Program of the NIH, National Institute of Environmental Health Sciences.

\setcounter{equation}{0}
\renewcommand{\theequation}{S\arabic{equation}}
\renewcommand{\thetable}{S\arabic{table}}

\section*{SI Text}
This supplement contains four sections. The first section derives the expected value of probability that the minimum P-value corresponds to a true signal followed by a more general result for the expected proportion of genuine signals among $u$ smallest P-values. The second section explores the limiting behavior of the $\Pr(H_A \mid \text{minP})$ expectation, i.e., EPGS for the top hit, assuming a constant prior probability of a true association, $\Pr(H_A)$. In the third section, we discuss how the limiting behavior of $\Pr(H_A \mid \text{minP})$ may change under the assumption of a decreasing $\Pr(H_A)$ as a function of the number of tests, $k$. In the fourth section, we describe enrichment of top hits by genuine signals in terms of the effect size expectation.

\section*{S1. Expected proportion of genuine signals}

Let $\gamma$ denote a single parameter that captures deviation from the null hypothesis and governs power. For a normally distributed test statistic, $\gamma$ would be the normal mean, shifted away from zero. For chi-squared, $F$, and $t$ statistics, $\gamma$ would be the noncentrality parameter of the corresponding distribution. In all these cases, zero value of $\gamma$ corresponds to the point null hypothesis, i.e., $H_0: \gamma = 0$ versus the alternative hypothesis $H_A: \gamma>0$. The point null assumption is embedded in the computation of two-sided P-values, however we should not be restricted to zero effect size models when we define which signals are considered genuine. For example, the null set of effect sizes may be represented by effects that are not large enough to be of interest. Therefore, we may define a cutoff value, $\gamma^0$ which divides the effect size distribution, $\Gamma$ into two subsets, $\Gamma_0$, that corresponds to the null set of effects with hypothesis $H_0$ and $\Gamma_A$ that corresponds to $H_A$. A continuous effect size distribution can be approximated by a finite mixture with a large number of components, $t$. Then the marginal CDF of P-values can be be expressed as a weighted sum:
\begin{eqnarray}
  \label{sieq:marg}
   \tilde{F}(p) = \sum_{i=1}^t w_i \,\, F(p \mid \gamma_i),
\end{eqnarray}
for some weights $w_1, \ldots, w_t$. Next, we partition $t$ effect sizes into $\Gamma_0:(\gamma_1, \ldots, \gamma_o)$ and $\Gamma_A: (\gamma_{o+1}, \ldots, \gamma_t)$ and define the null hypothesis as $H_0: \max(\gamma) \leq \gamma_o$. To derive the expected probability that a finding is genuine, we let 
\begin{equation*}
  T =
  \begin{cases*}
   1 & if $\max(\gamma) > \gamma_o$, \\
   0 & otherwise.
  \end{cases*}
\end{equation*}
Further, we assume that $k=L+m$ tests were performed with $\left\{P_{\Gamma_0} \right\}$ being a set of P-values generated by signals from $\Gamma_0:(\gamma_1, \ldots, \gamma_o)$ and $\left\{P_{\Gamma_A} \right\}$ being a set of P-values generated from $\Gamma_A:(\gamma_{o+1}, \ldots, \gamma_t)$. We use the notation $P_{\Gamma_A}^{(j:m)}$ to denote $j$-th smallest P-value out of $m$ in total, that originated from an effect in the set $\Gamma_A$. We apply a similar notation to the distribution functions, e.g. $\tilde{F}_{\Gamma_A}^{(j:m)}$ denote the marginal CDF of $j$-th ordered P-value among $m$ in total.

Next, we consider the expectation of the probability that $T$ is 1. For the minimum P-value,
\begin{eqnarray*}
 \text{EPGS} &=& E \left[ \Pr(H_A \mid \text{minP})\right] \\ 
             & = &  E\left[\Pr(T=1)\right] \\ \nonumber
             & = & \Pr \left\{ \min\left\{P_{\Gamma_A} \right\} < \min\left\{P_{\Gamma_0} \right\}\right\} \\ \nonumber 
             & = & \int_0^1 \int_0^{q^{(1:L)}(p)} \tilde{f}_{{\Gamma}_A}(t)dt\; dp,
\end{eqnarray*}
where $q^{(1:L)}(\cdot)$ is a function such that $q^{(1:L)}(1 - \tilde{F}_{\Gamma_0}^{(1:L)}(x)) = x$ with $\tilde{F}^{(1:L)}_{\Gamma_0}(\cdot)$ being the  marginal CDF for the minimum P-value that originated from a signal of the set $\Gamma_0$, i.e., $\tilde{F}_{\Gamma_0}^{(1:L)}(x) = 1 - \left[1 - \tilde{F}_{\Gamma_0}(x)\right]^L$ and $\tilde{f}_{{\Gamma}_A}(\cdot)$ is the marginal PDF of P-values over $\Gamma_A: (\gamma_{o+1}, \ldots, \gamma_t)$. Further, 
\begin{eqnarray*}
  \int_0^1 \int_0^{q^{(1:L)}(p)} \tilde{f}_{{\Gamma}_A}(t)dt\; dp & = & \int_0^1 \tilde{F}_{{\Gamma}_A}^{(1:m)} (\tilde{Q}_{{\Gamma}_0}^{(1:L)}(U)) dp \\ \nonumber
 & = & E \left[\tilde{F}_{{\Gamma}_A}^{(1:m)} (\tilde{Q}_{{\Gamma}_0}^{(1:L)}(U)) \right],
\end{eqnarray*}
where $Q_{{\Gamma}_0}^{(1:L)}(\cdot)$ is a quantile function, i.e., $Q_{{\Gamma}_0}^{(1:L)}(F_{{\Gamma}_0}^{(1:L)}(x)) = x$, and $U$ is a uniform (0,1) random variable.

Extending this to the higher order statistics, we define
\begin{eqnarray}
   \Pca_{j,u} &=& \Pr \left\{ P^{(j:m)}_{\Gamma_A} < P^{(u-j+1:L)}_{\Gamma_0} \right\} \nonumber \\
    &=& E \left[\tilde{F}_{{\Gamma}_A}^{(j:m)} (\tilde{Q}_{{\Gamma}_0}^{(u-j+1:L)}(U)) \right] \quad (j \le u).
\label{rankprob}
\end{eqnarray}
It is worth to note that $\Pca_{j,u}$ is simply an expectation of a function of a uniform random variable and thus it is straightforward to evaluate numerically.
 
Finally, the $\Pca_{j,u}$ values from  Eq. \ref{rankprob} define the distribution of ordered P-values in the set $\Gamma_A$. Specifically, $\Pca_{j,u}$ is the expected chance that at least $j$ genuine signals will rank among top $u$ smallest P-values, therefore the expected proportion of genuine signals among $u$ smallest P-values is given by their average:
\begin{eqnarray}
  \label{sieq:nnull}
  \text{EPGS} & = & \frac{1}{u} \sum_{j=1}^{\min(m,u)} \Pca_{j,u}.
\end{eqnarray}

To validate our derivation, we compared theoretical values based on Eq. \ref{rankprob} to empirical expectations computed based on $B$=1,000,000 simulations. Table \ref{sitab1} summarizes results of several comparisons, indicating agreement between the theoretical and the empirical expectations. For the table, we took several random values of $k$ and $m$. Effect sizes were taken to be equally spaced between zero and one, $0 < \gamma_1, \ldots, \gamma_t \le 1$. Weights for these effect sizes were obtained from a discretized Gamma distribution, Gamma(shape, 1/shape), where the shape was randomly chosen to fall between 1/4 and 4 for each row in the table. Empirical values were obtained by assuming normally distributed statistics with noncentralities $\gamma_1, \ldots, \gamma_t$, where the boundary between the null, $\Gamma_0$, and the alternative set, $\Gamma_A$ was placed at the median value. After obtaining $L+m=k$ statistics in each simulation, the values were sorted and the count $C$ incremented if $j$-th ordered P-value originated from the set $\Gamma_A$ was among $u$ smallest P-values. The empirical estimate of $\Pca_{j,u}$ is $C/B$.
\begin{table}[h]
  \centering
  \caption{Comparison of theoretical and empirical values of $\Pca_{j,u}$.}
  \begin{tabular}{ccccccc}
  j & u & m  & k & Shape & Empirical $\Pca_{j,u}$ & Theoretical $\Pca_{j,u}$ \\
    \midrule
   1  &  1  &  53  &  179  &  1.19  &  0.603  &  0.603 \\
   1  &  1  &  42  &  142  &  1.73  &  0.593  &  0.593 \\
   2  &  14  &  11  &  122  &  0.34  &  0.766  &  0.767 \\
   2  &  26  &  10  &  263  &  0.35  &  0.621  &  0.621 \\
   3  &  26  &  14  &  219  &  1.55  &  0.750  &  0.751 \\
   3  &  16  &  14  &  165  &  2.14  &  0.667  &  0.667 \\
   4  &  6  &  17  &  111  &  0.46  &  0.404  &  0.404 \\ 
   4  &  12  &  22  &  106  &  0.4  &  0.909  &  0.910 \\ 
   5  &  13  &  19  &  111  &  3.87  &  0.768  &  0.768 \\
   5  &  25  &  15  &  168  &  4.00  &  0.595  &  0.596 \\
   6  &  9  &  33  &  166  &  0.51  &  0.671  &  0.672 \\ 
   6  &  14  &  42  &  176  &  0.35  &  0.946  &  0.946 \\ 
   7  &  11  &  40  &  191  &  0.53  &  0.767  &  0.766 \\
   7  &  9  &  35  &  150  &  3.07  &  0.771  &  0.771 \\
   8  &  17  &  27  &  268  &  0.97  &  0.197  &  0.197 \\
   8  &  11  &  32  &  159  &  0.32  &  0.547  &  0.547 \\
   9  &  17  &  38  &  222  &  0.62  &  0.656  &  0.656 \\
   9  &  13  &  19  &  157  &  2.68  &  0.080  &  0.080 \\
   10  &  30  &  34  &  292  &  0.26  &  0.373  &  0.373 \\
   10  &  22  &  35  &  252  &  0.85  &  0.496  &  0.496 \\
   11  &  15  &  44  &  268  &  0.58  &  0.384  &  0.383 \\
   11  &  13  &  34  &  199  &  0.26  &  0.195  &  0.195 \\
   12  &  13  &  63  &  291  &  2.43  &  0.696  &  0.695 \\ 
   12  &  17  &  54  &  278  &  0.44  &  0.617  &  0.617 \\
    \bottomrule
  \end{tabular}
  \label{sitab1}
\end{table}

\section*{S2. Limiting behavior of  the expected probability that a finding is genuine with constant prior probability of a true association}

In this section, we describe a limiting behavior of the $E \left[ \Pr(H_A \mid \text{minP})\right]$ as the number of tests, $k$, goes to infinity. In our previous work we considered a special case of the results from the previous section, by setting the effect size under $H_0$ to zero, i.e., by assuming the point null hypothesis\cite{kuo2011novel}. This simplification leads to approximations that are useful for studying dependency of EPGS on the number of tests. Let $\pi = \Pr(H_0)$ be the prior probability of the null hypothesis and $1-\pi = \Pr(H_A)$ be the probability of the alternative hypothesis. Let $G_0(\cdot)$ and $G_\gamma(\cdot)$ denote the cumulative distribution function (CDF) of the test statistic under the null and the alternative hypothesis respectively. The CDF of P-value derived from the continuous test statistics where the deviation from the null can be described by a single parameter $\gamma$ can be written as
 \begin{eqnarray}
   \label{sieq:1}
    F_\gamma(p) = 1 - G_\gamma\left[G_0^{-1}(1-p) \right] .
 \end{eqnarray}
 Assuming that $\gamma$ follows the distribution $\Gamma(\gamma)$ and averaging over all possible values of the effect size for genuine signals, the marginal CDF of a true association P-value is:
 \begin{eqnarray}
   \label{sieq:2}
    \tilde{F}(p) = \int F_\gamma(p)\Gamma(\gamma)d\gamma.
 \end{eqnarray}
 Futher, the CDF of the $j$th-ordered true association P-value has the form:
 \begin{eqnarray}
   \label{sieq:3}
   \tilde{F}^{(j:m)} = 1 - \text{Bin}(j-1; \,\, m, \tilde{F}(p)),
 \end{eqnarray}
 where Bin denotes the binomial CDF evaluated at $j-1$ successes in $m$ true associations with success probability $\tilde{F}(p)$. Assuming that the effect size is zero under $H_0$, the expected proportion of genuine signals among $u$ smallest out of $k$ total sorted list of P-values, $\{p_{(1)}, \ldots p_{(k)}\}$
\begin{eqnarray}
   \label{sieq:pgs}
   \text{EPGS} \approx \frac{1}{u}\sum_{j=1}^{\min(u,m)}\tilde{F}^{(j:m)}\left(\frac{u - j + 1}{(\pi \cdot k +1)(1 + 1/u)} \right).
\end{eqnarray}

In this section, we focus on the limiting behavior of EPGS for $u = 1$ (i.e., the probability that one of the true associations will have the smallest P-value in a study) under the assumption that the rate of occurrence of true signals does not depend on $k$. Considering minP, we define the expected false discovery rate (EFDR) as EFDR = 1-EPGS. Then
 \begin{eqnarray}
   \label{sieq:lnFDR} 
   \ln\{ \text{EFDR}\} \approx -(1-\pi)k \cdot \tilde{F}\left( \frac{1}{\pi k} \right) = - \frac{\tilde{F}\left( \frac{1}{\pi k} \right)}{1/((1-\pi)k)}.
 \end{eqnarray}
 Consider the limit of the expression in Eq. \ref{sieq:lnFDR} as $k \rightarrow \infty$ 
 \begin{eqnarray*}
    \lim_{k\to\infty}- \frac{\tilde{F}\left( \frac{1}{\pi k} \right)}{1/((1-\pi) k)} & = & - \lim_{k\to\infty} \frac{-\tilde{f}\left( \frac{1}{\pi k} \right)/(k^2 \pi)}{-1/(k^2 (1-\pi))} \\
& = & -\lim_{k\to\infty} \tilde{f}\left( \frac{1}{\pi k} \right) \cdot \frac{1-\pi}{\pi} \\
& = & - \infty; \quad \forall \gamma>0.
 \end{eqnarray*}
and EPGS $\to 1$ At $\gamma=0$, EFDR $\approx \exp(-\pi/(1-\pi)) \approx 1-\pi =\Pr(H_0)$, as it should be, since the distinction between genuine and false signals becomes merely a label.

\section*{S3. Decreasing prior probability of a true association}
In this section, we discuss how the limiting behavior of  EPGS=$E \left[ \Pr(H_A \mid \text{minP})\right]$ changes if $\Pr(H_A)$ is a decreasing function of $k$, $1-\pi = h(k)$, with $$ \lim_{k \to \infty}h(k) = 0.$$
From Eq. \ref{sieq:lnFDR}, it follows that
\begin{eqnarray}
   \lim_{k\to\infty}(\ln\{ \text{EFDR}\}) &\approx& 
\lim_{k\to\infty} -h(k)k \cdot \tilde{F}\left( \frac{1}{(1-h(k))k}  \right). \nonumber \\
% & = & \lim_{k\to\infty} - \frac{F\left( 1/\left[(1-h(k))k\right]  \right)}{1/(h(k) k)}.
\label{sieq:lim}
\end{eqnarray}
Regardless of the specific form of $h(k)$, $\tilde{F}\left[1/(k(1-h(k)))\right] \to 0$ in Eq. \ref{sieq:lim} as $k$ approaches infinity. Therefore, if 
\begin{eqnarray}
  \label{sieq:hk}
  \lim_{k \to \infty}h(k)\cdot k = c,
\end{eqnarray}
where is a positive constant, then $E \left[ \Pr(H_A \mid \text{minP})\right]\rightarrow 0$ as $k \rightarrow \infty$. For example, $(1 - \pi^{1/k})\cdot k$ initially increases with $k$, but reaches a constant, $\lim_{k \to \infty}(1 - \pi^{1/k})\cdot k  = -\ln(\pi) > 0$.

If $\lim_{k \to \infty}h(k)\cdot k = \infty$, the limit in Eq. \ref{sieq:lim} may or may not go to infinity, depending on the steepness of decrease in the rate of $\Pr(H_A)$. To examine specific cases we make further simplifications: by noting that EPGS increases with $\gamma$, we consider a lower bound on the effect size and assume a single value $\gamma>0$. One example of a slow decreasing rate is a logarithmic decrease in the rate of true associations with $k$,
\begin{eqnarray*}
   \lim_{k\to\infty}(\ln\{ \text{EFDR}\}) &\approx& \lim_{k\to\infty} -h(k)k \cdot \tilde{F}\left( \frac{1}{(1-h(k))k}  \right) \\ \nonumber
                                         & = & \lim_{k\to\infty} - \frac{\tilde{F}\left( 1/\left[(1-h(k))k\right]  \right)}{1/(h(k) k)}. \nonumber
\end{eqnarray*}
Futher,
\begin{eqnarray*}
  \left[ \frac{1}{h(k)k} \right]' =  \frac{1}{k^2 (1-\pi)} - \frac{\ln(k)}{k^2(1-\pi)}; \\ \nonumber
\end{eqnarray*}
and
\begin{eqnarray*}
&&  \tilde{F}'\left( \frac{1}{(1-h(k))k}  \right)  \\
& = & \left[ \frac{1}{k^2(1-\frac{1-\pi}{ \ln(k)})} + \frac{1-\pi}{k^2(1-\frac{1-\pi}{ \ln(k)})^2 \ln^2(k)} \right] \\
 && \times \tilde{f}\left( \frac{1}{(1-h(k))k} \right).         
\end{eqnarray*}
Accordingly, by L'Hospital's Rule, $\lim_{k\to\infty}(\ln\{ \text{EFDR}\})$ in Eq. \ref{sieq:lim} is
\begin{eqnarray}
  && \lim_{k\to\infty} - \frac{(1-\pi)(1-\pi + (1-\pi - \ln(k))) \ln(k)}{(1-\pi - \ln(k))^2 (\ln(k) - 1)} \\ \nonumber
  && \times \tilde{f}\left( \frac{1}{(1-h(k))k} \right). \\
\label{sieq:ln}
\end{eqnarray}
The first term of the product in Eq. \ref{sieq:ln} goes to zero as $k \rightarrow \infty$ but the second term, $\tilde{f}\left( \frac{1}{(1-h(k))k} \right)$, approaches infinity faster, assuming $k$ tests were based on a normal statistic (computed with Wolfram Mathematica \cite{wolfram2008wolfram}). 

As in the case of constant $\Pr(H_A)$, $\ln\{\text{EFDR}\} \rightarrow -\infty$ implies that $\text{EPGS} \to 1$. Therefore, even if the probability of finding a causal variant decreases logarithmically with the number of variants considered, the probability that the smallest P-values is a true finding still approaches 1 as more tests are performed.

\section*{S4. Expected effect size of ordered  P-values}
\begin{table}[h]
  \centering
  \caption{Comparison of theoretical and empirical values of $E \left(\gamma \mid X^{(i:k)} \right)$ allowing for zero effect sizes.}
\begin{tabular}{|c|c|c|c|c|c|c|}
\hline 
\multicolumn{7}{|c|}{Allow point $H_0$: Pr(min($\gamma$)=0)=0.95}\\
\hline 
\hline 
$i\Rightarrow$ & 1 & 25 & 1 & 25 & 1 & 25 \\
\hline
$k\Downarrow$ & \multicolumn{2}{c|}{Simulated $E(\gamma)$} & \multicolumn{2}{c|}{Analytical $E(\gamma)$} & \multicolumn{2}{c|}{$E(\chi^2)$}\\
\hline 
1K   & 1.52 & 0.28  & 1.53 & 0.27  & 14.00 & 5.45  \\
5K   & 2.28 & 0.64  & 2.30 & 0.63  & 18.20 & 8.78  \\
10K  & 2.51 & 0.86  & 2.60 & 0.87  & 20.09 & 10.34 \\
100K & 3.31 & 1.92  & 3.36 & 1.91  & 26.57 & 16.04  \\
150K & 3.42 & 2.14  & 3.46 & 2.11  & 27.73 & 17.13  \\
200K & 3.47 & 2.27  & 3.52 & 2.24  & 28.55 & 17.90  \\
%250K & 3.50 & 2.33  & 3.57 & 2.35  & 29.18 & 18.51  \\
350K & 3.59 & 2.45  & 3.64 & 2.50  & 30.14 & 19.44  \\
400K & 3.67 & 2.54  & 3.66 & 2.55  & 30.52 & 19.81  \\
500K & 3.68 & 2.67  & 3.70 & 2.65  & 31.16 & 20.43  \\
550K & 3.69 & 2.67  & 3.72 & 2.69  & 31.43 & 20.70  \\
650K & 3.73 & 2.74  & 3.75 & 2.75  & 31.90 & 21.17  \\
\hline
\hline
\multicolumn{7}{|c|}{Thresholding: minP$>$0.05/$k$; \, Allow point $H_0$: Pr(min($\gamma$)=0)=0.95}\\
\hline 
\hline 
$i\Rightarrow$ & 1 & 25 & 1 & 25 & 1 & 25 \\
\hline
$k\Downarrow$ & \multicolumn{2}{c|}{Simulated $E(\gamma)$} & \multicolumn{2}{c|}{Analytical $E(\gamma)$} & \multicolumn{2}{c|}{$E(\chi^2)$}\\
\hline 
1K   & 1.30 & 0.27  & 1.29 & 0.27 &  12.75 & 5.43  \\
5K   & 1.99 & 0.63  & 1.98 & 0.63 &  16.43 & 8.75  \\
10K  & 2.28 & 0.86  & 2.27 & 0.86 &  18.05 & 10.30 \\
100K & 3.00 & 1.89  & 3.04 & 1.89 &  23.43 & 15.95 \\
150K & 3.15 & 2.10  & 3.14 & 2.09 &  24.36 & 17.01 \\
200K & 3.23 & 2.24  & 3.21 & 2.22 &  25.02 & 17.78 \\
%250K & 3.26 & 2.33  & 3.26 & 2.33 &  25.53 & 18.38 \\
350K & 3.33 & 2.49  & 3.33 & 2.47 &  26.29 & 19.29 \\
400K & 3.34 & 2.50  & 3.36 & 2.53 &  26.59 & 19.65 \\
500K & 3.41 & 2.63  & 3.40 & 2.62 &  27.09 & 20.26 \\
550K & 3.42 & 2.63  & 3.42 & 2.66 &  27.30 & 20.52 \\
650K & 3.44 & 2.73  & 3.45 & 2.73 &  27.67 & 20.98 \\
\hline
\end{tabular}
\label{sitab2}
\end{table}
%%%%%%

\begin{table}[h]
  \centering
  \caption{Comparison of theoretical and empirical values of $E \left(\gamma \mid X^{(i:k)} \right)$ without allowing for zero effect sizes.}
\begin{tabular}{|c|c|c|c|c|c|c|}
\hline
\multicolumn{7}{|c|}{Pr(min($\gamma$)=0.05)=0.95}\\
\hline
\hline
$i\Rightarrow$ & 1 & 25 & 1 & 25 & 1 & 25 \\
\hline
$k\Downarrow$ & \multicolumn{2}{c|}{Simulated $E(\gamma)$} & \multicolumn{2}{c|}{Analytical $E(\gamma)$} & \multicolumn{2}{c|}{$E(\chi^2)$}\\
\hline 
1K  & 1.42  & 0.30  & 1.40 & 0.31  & 14.28 & 5.68  \\
5K  & 2.09  & 0.63  & 2.12 & 0.63  & 18.43 & 9.08  \\
10K & 2.39  & 0.82  & 2.42 & 0.83  & 20.29 & 10.65 \\
100K & 3.16 & 1.75  & 3.23 & 1.76  & 26.67 & 16.31  \\
150K & 3.27 & 1.95  & 3.34 & 1.94  & 27.81 & 17.37  \\
200K & 3.37 & 2.08  & 3.41 & 2.07  & 28.62 & 18.13  \\
%250K & 3.42 & 2.17  & 3.47 & 2.17  & 29.25 & 18.73  \\
350K & 3.50 & 2.31  & 3.54 & 2.32  & 30.20 & 19.64  \\
400K & 3.55 & 2.37  & 3.57 & 2.37  & 30.58 & 20.00  \\
500K & 3.59 & 2.47  & 3.62 & 2.47  & 31.21 & 20.61  \\
550K & 3.61 & 2.48  & 3.64 & 2.51  & 31.48 & 20.87  \\
650K & 3.64 & 2.58  & 3.67 & 2.57  & 31.95 & 21.33  \\
\hline
\hline 
\multicolumn{7}{|c|}{Thresholding: minP$>$0.05/$k$; \, Pr(min($\gamma$)=0.05)=0.95}\\
\hline 
\hline 
$i\Rightarrow$ & 1 & 25 & 1 & 25 & 1 & 25 \\
\hline
$k\Downarrow$ & \multicolumn{2}{c|}{Simulated $E(\gamma)$} & \multicolumn{2}{c|}{Analytical $E(\gamma)$} & \multicolumn{2}{c|}{$E(\chi^2)$}\\
\hline 
1K   & 1.19 & 0.28 &  1.19 & 0.31  & 12.98 & 5.66  \\
5K   & 1.80 & 0.64 &  1.81 & 0.62  & 16.61 & 9.05  \\
10K  & 2.10 & 0.83 &  2.08 & 0.82  & 18.20 & 10.61 \\
100K & 2.88 & 1.76 &  2.87 & 1.74  & 23.50 & 16.21  \\
150K & 2.95 & 1.91 &  2.98 & 1.92  & 24.42 & 17.26  \\
200K & 3.07 & 2.05 &  3.06 & 2.05  & 25.07 & 18.01  \\
%250K & 3.12 & 2.15 &  3.12 & 2.15  & 25.58 & 18.60  \\
350K & 3.23 & 2.31 &  3.20 & 2.29  & 26.33 & 19.49  \\
400K & 3.23 & 2.34 &  3.23 & 2.35  & 26.63 & 19.85  \\
500K & 3.26 & 2.46 &  3.28 & 2.44  & 27.13 & 20.44  \\
550K & 3.30 & 2.47 &  3.30 & 2.48  & 27.34 & 20.70  \\
650K & 3.32 & 2.52 &  3.33 & 2.55  & 27.71 & 21.15  \\
\hline
\end{tabular}
\label{sitab3}
\end{table}

In preceding sections, we adopted a framework that makes distinction between effect sizes that are large enough to be considered genuine and correspond to $H_A$ and a set of smaller effect sizes that correspond to $H_0$. Such dichotomization allowed us to directly obtain expected enrichment of a set of $u$ smallest P-values by genuine signals as a function of the number of tests. A conceptually different approach is to model all $k$ tested signals as arising from a single effect size distribution.  This approach allows for thresholding, for example, Fig. \ref{fig3} of the main text presents the enrichment of top hits by genuine signals across experiments that are subject to thresholding by a significance cutoff: the average was taken only across nonsignificant experiments, i.e., those where minP did not reach the Bonferroni threshold (minP$>0.05/k$). Continuing the finite mixture approach (Eq. \ref{sieq:marg} and further), let $X^{(i:k)}$ denote the test statistic that corresponds to the $j$-th smallest P-value, $P^{(j:k)}$. In terms of the statistics, $X^{(i:k)}$ is the $i$th largest value, $i=k-j+1$. For the thresholding process just described, we are interested if the enrichment still takes place among nonsignificant multiple testing experiments, thus we also consider the expectation for statistics for which the respective P-values did not reach the critical value $\alpha/k$. For positively valued statistics, such as the chi-squared, and allowing for thresholding, the expected value can be derived as
\begin{eqnarray}
&&   E \left( X^{(i:k)} \right) = \nonumber \\
&& \int_0^{\tilde{F}(Q(1-\frac{\alpha}{k}))} \text{Bin}\left\{i-1 \mid k, \frac{\tilde{F}(s)}{\tilde{F}(Q(1-\frac{\alpha}{k}))} \right\} ds \nonumber \\
\label{siexpx}
\end{eqnarray}
where $Q(\cdot)$ is the inverse CDF of the test statistic. The expectation for the effect sizes that correspond to this $i$th largest and possibly thresholded statistic can be well approximated as
\begin{eqnarray}
  E \left(\gamma \mid X^{(i:k)} \right) &\approx& 
    \frac{ \sum_i^t \gamma_i \, w_i \, f_{\gamma_i}\left\{E\left(X^{(i:k)} \right)\right\}}
    {\tilde{f}\left\{E\left(X^{(i:k)}\right)\right\} } \label{siexpgamma}
\end{eqnarray}
For a fixed value of $i$, both expectations are increasing with $k$ For the statistic in Eq. \ref{siexpx}, this can be shown simply by applying the monotone transformation $U=1-\tilde{F}(\cdot)$ to ordered values of $X$. Then $U^{(i:k)}$ is the $i$th smallest value, $E(U^{(i:k)}=i/(k+1)$ is decreasing with $k$, therefore the expectation in \ref{siexpx} is increasing with $k$. The expectation in Eq. \ref{siexpgamma} is an increasing function of $E \left( X^{(i:k)} \right)$. This can be seen by examining the sum in this equation as being taken over an ordered sequence of noncentral densities. As the argument of $x$ of $\tilde{f}_{\gamma_i}(x)$ increases, densities indexed by small noncentralities $\gamma_i$ contribute increasingly smaller values to the sum. Averages over $B$=10,000 simulation experiments, designed similarly to those used to produce Table \ref{sitab1}, reveal that the approximation in Eq \ref{siexpgamma} is very good. These results are given in Tables \ref{sitab2}, \ref{sitab3}. Allowing for zero effect sizes in Table \ref{sitab2} was achieved by setting the smallest of ordered $\gamma_1,\dots,\gamma_t$, to zero, that is, $\gamma_1=0$, and the notation Pr(min($\gamma$)=0)=0.95 reflects that the corresponding mixture weight was $w_1=0.95$.

\bibliography{MinP.Arxiv.9.6.16.bib}
\end{document}